\documentstyle[twoside,fleqn,espcrc2,epsfig,psfig]{article}

\newcommand{\AmS}{{\protect\the\textfont2

  A\kern-.1667em\lower.5ex\hbox{M}\kern-.125emS}}

\title{Cluster Percolation and Explicit Symmetry Breaking in Spin Models
\thanks{The work has been supported by the TMR network ERBFMRX-CT-970122
and the DFG under grant FOR 339/1-2.} 
}

\author{S. Fortunato, H. Satz
\\\vskip 6pt
Fakult{\"a}t f{\"u}r Physik, 
    Universit{\"a}t Bielefeld,\\
    D-33501 Bielefeld, Germany} 
 
\begin{document}

\begin{abstract}

Many features of spin models can be interpreted
in geometrical terms by means of the properties of well
defined clusters of spins. In case of spontaneous
symmetry breaking, the phase transition of models like the q-state Potts model, 
$O(n)$, etc., can be
equivalently described as a percolation transition of clusters.
We study here the behaviour of 
such clusters when the presence of an external field $H$
breaks explicitly the global symmetry of the Hamiltonian of the theory.
We find that these clusters 
have still some interesting relationships with thermal
features of the model. 

\end{abstract}

\maketitle

\section{INTRODUCTION}

Since the early 40's \cite{onsager} the interest of several physicists studying
critical phenomena has focused on the 
investigation of geometrical structures of the system whose interplay
could represent the mechanism for the occurrence of a phase transition.

A breakthrough in this course of studies was the discovery of
Fortuin and Kasteleyn (FK) \cite{FK} that the $q$-state Potts model
with vanishing external field
can be reformulated as a geometrical model, with the spin configurations
turning into clusters configurations by connecting nearest-neighbouring 
spins of the same orientation with a
bond probability $p_B=1-\exp(-J/kT)$ ($J$ is the Potts coupling, $T$ the temperature).
A remarkable consequence of this
mapping is that the FK clusters percolate at the critical point of the thermal transition
and, in case of second order phase transitions, the percolation 
exponents coincide with the thermal exponents. This result has been since then
extended to a wide variety of models, including 
continuous spin models like $O(n)$ \cite{on}. 

The absence of an external field endows the Hamiltonian of all
these models with a global symmetry, which is spontaneously broken
by the state of the system at low temperatures. The spontaneous breaking
of such global symmetry is, ultimately, the reason of the phase transition.
If we switch on an external field $H$, the Hamiltonian of the system
breaks explicitly the global symmetry one has for $H=0$, and this has
deep consequences on the possibility for the model to
show critical behaviour. One has, in general, two possible cases, 
according to the order of the phase transition when there is no external field:

\begin{itemize}
\item{The model undergoes a second order phase transition for $H=0$.} 
\item{The model undergoes a first order phase transition for $H=0$.}
\end{itemize}

Here we want to see whether the clusters that describe the 
critical behaviour at $H=0$ have any relationship
with thermal properties of the model also in presence of an 
external field. For this purpose we will analyse
separately the two cases listed above. 

\section{SECOND ORDER PHASE TRANSITION FOR H=0}

If the system undergoes a second order phase transition for $H=0$, 
for any $H{\neq}0$ the partition function of the system is not singular
and one has at most a rapid crossover from the ordered to the disordered
phase. 

From the renormalization group ansatz of the magnetization as a function 
of $h=H/kT$ and $t=(T-T_c)/T_c$ one derives  
simple scaling laws for $t,h{\ll}1$. Particularly interesting is the 
so called {\it pseudocritical line}, which is given by the 
temperature $t_{\chi}$ at which the susceptibility $\chi$ peaks 
for a given $h$. The equation of the pseudocritical line is given by
\begin{eqnarray}
t_{\chi}\,\propto\,h^{1/\beta\delta},
\label{uno}
\end{eqnarray}
where $\beta$ and $\delta$ are critical exponents.
\vskip-0.7cm
\begin{figure}[htb]
\epsfig{file=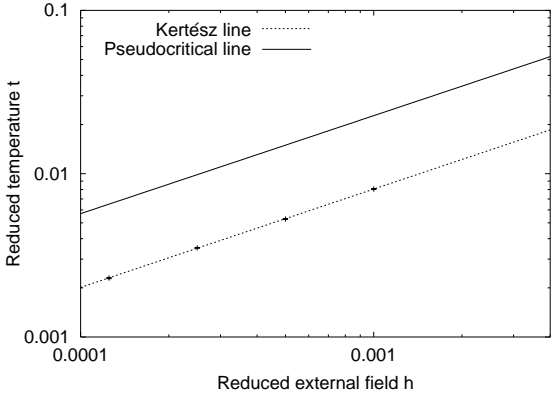,width=75mm}
\vspace{-12mm}
\caption{Comparison of the Kert\'esz and the pseudocritical line
for the 2D Ising model.}
\label{kert}
\end{figure}
\vskip-0.3cm
On the other hand, the FK clusters\footnote{The result
is valid for any model which can be mapped onto
a cluster model, e.g. O(n).} percolate for any $H{\neq}0$: the 
corresponding geometrical variables show divergences as in the case
$H=0$ and percolation remains a genuine critical phenomenon. 
The line of percolation
thresholds is called {\it Kert\'esz line} \cite{kertesz}.

We want to compare the Kert\'esz and the pseudocritical line.
Fig. \ref{kert} shows the two lines for the 2D Ising model. In the 
double logarithmic scale of the plot, they look parallel
to each other. A fit of the four points of the percolation line
leads indeed to the same scaling law as Eq. \ref{uno}; the relative exponent
is $\kappa=0.534(3)$, in excellent agreement with $1/\beta\delta=8/15=0.533$.
We have also found that the same result holds for the 3D O(2) model \cite{san},
so that it is likely to be general.

\section{FIRST ORDER PHASE TRANSITION FOR H=0}

In the case of a first order transition for $H=0$, the model 
shows a discontinuous phase change also in the presence of an external
field, as long as $H$ is smaller than some critical $H_c$. 
One has then a line of first order phase transition thresholds, from
$H=0$, $T=T_c(0)$ to
$H=H_c$ and $T=T_c(H_c)$ ({\it endpoint}), where
the transition
becomes continuous and the exponents are conjectured to be the Ising exponents.

We have examined the case of the 3D 3-state Potts model \cite{ios}, 
which is subject of intense investigations
because its phase transition is closely
related to the deconfinement transition of finite temperature QCD
\cite{S&Y}. The critical value of the field $H_c$ 
was recently
determined with great precision \cite{sven}.
\vskip-0.7cm
\begin{figure}[htb]
\epsfig{file=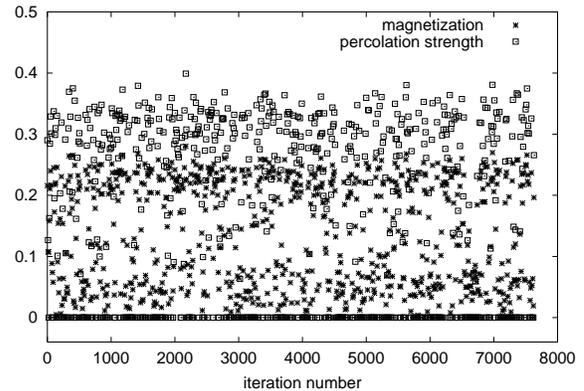,width=75mm}
\vspace{-12mm}
\caption{Time history  
of $M$ and $P$ at a point of the first order transitions line. The lattice size is $70^3$.}
\label{his}
\end{figure}
\vskip-0.3cm
In order to see how the FK clusters behave for $H{\neq}0$ we 
have analyzed the time history of the percolation order parameter,
the percolation strength $P$, which is the probability that
a randomly chosen site of the lattice belongs to a percolating
cluster. Fig. \ref{his} shows the time history of $P$ compared to 
the one of the magnetization
$M$ at a point of the first order phase transitions line, quite close to
the endpoint. From the figure we can see that $P$
makes a jump from zero to a non-zero value: there is then 
a discontinuous geometrical transition between a percolation and a non-percolation phase. 
We repeated the procedure for several points of the first order transitions
line, obtaining any time the same result.
 
\begin{figure}[htb]
\epsfig{file=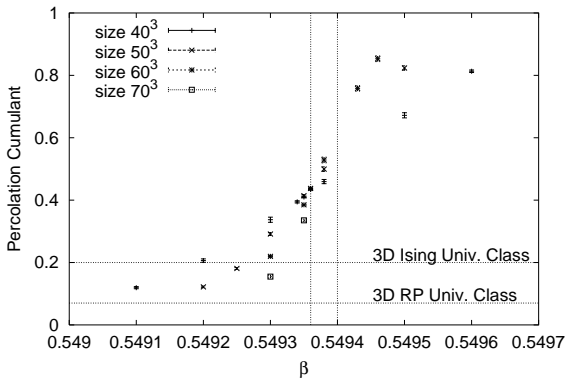,width=75mm}
\vspace{-12mm}
\caption{Percolation cumulant at
$H=H_c$ for FK clusters as a function of $\beta$ for four 
lattice sizes.}
\label{end}
\end{figure}

For $H=H_c$ we expect that the percolation transition of the FK clusters becomes
continuous like the thermal one. It is then interesting to check whether
the two thresholds coincide and, in this case, whether the percolation
exponents belong to the 3D Ising universality class as the thermal exponents. 
Fig. \ref{end} shows the percolation cumulant, that is the
fraction of configurations with a percolating cluster, as a function
of $\beta=J/kT$ for different lattice sizes. The crossing point of the curves 
is the critical point of the geometrical transition and it is in good
agreement with the thermal threshold determined in \cite{sven}
(vertical dashed lines in the figure).
\vskip-0.7cm
\vskip-0.3cm
The height of the crossing point is a universal number: the figure shows that it 
considerably differs from the values corresponding to the 3D Ising and random percolation
(RP) universality classes (horizontal lines in the figure).
A finite size scaling analysis of the cluster variables at the 
critical point leads to the following values for the percolation
exponents: $\beta/\nu=0.32(3)$,
$\gamma/\nu=2.32(2)$, $\nu=0.45(3)$. 
As expected, these exponents belong neither to the 3D Ising nor to the RP universality class.

\section{CONCLUSIONS}

We have seen that the clusters which describe the critical behaviour
of thermal models in absence of an external
field $H$ maintain a close relationship with thermal features of the models
also when $H{\neq}0$. 

If the spontaneous symmetry breaking transition is second order, in the limit of 
small fields the Kert\'esz line is described by the same function as 
the pseudocritical line of the model, a power law with exponent $1/\beta\delta$.

If the spontaneous symmetry breaking transition is first order,
the line of thermal first order phase transitions is also a line of first order
percolation transitions for the clusters. In contrast to the 
magnetization, the percolation strenght $P$ is a good order 
parameter for the geometrical transition all along the line. The endpoint
is a percolation point, but the corresponding critical 
exponents are neither in the Ising nor in the random percolation
universality class. 
That suggests that one might need to change the cluster definition, eventually
introducing an explicit field dependence in the bond probability $p_B$, so to obtain
the correct critical exponents of the thermal transition at the endpoint.
Such definition could turn out to be fruitful also in the crossover region.

\end{document}